\documentstyle[aps,twocolumn,epsf]{revtex}

\newcommand{\cuti}      {$^{63}\!(T_1T)^{-1}$}
\newcommand{\ybcod}     {$\rm Y Ba_2 Cu_3 O_{7-\delta}$}
\newcommand{\tc}        {$T_c$}

\title{\bf Magnetic field independence of the spin gap in \mbox{$\bf
YBa_2Cu_3O_{7-\delta}$}}

\author{K. Gorny,$^{1}$ O. M. Vyaselev,$^{1}$\protect\cite{prbdd} J.
A. Martindale,$^{1}$ V.
A. Nandor,$^{1}$ C. H. Pennington,$^{1}$ P. C. Hammel,$^{2}$ W.
L. Hults,$^{2}$
J. L. Smith,$^{2}$ P. L. Kuhns,$^{3}$ A. P. Reyes,$^{3}$ and W. G.
Moulton$^{3}$}

\address{$^{1}$Department of Physics, Ohio State University, 174 W.
18th Ave. Columbus, OH 43210}
\address{$^{2}$Los Alamos National Laboratory, Los Alamos, New Mexico
87545}
\address{$^{3}$National High Magnetic Field Laboratory; Tallahassee,
FL 32310}

\date{Received: \quad September 15, 1998}

\address{
\parbox{14cm}{\bigskip\rm\small
We report, for magnetic fields of 0, 8.8, and 14.8 Tesla, measurements
of the temperature dependent $^{63}$Cu NMR spin lattice relaxation
rate
for near optimally doped \ybcod, near and above
$T_c$.
In sharp contrast with previous work we find no magnetic field
dependence.
We discuss experimental issues arising in measurements of
this required precision, and implications of the experiment regarding
issues including the spin or pseudo gap. 
}}

\begin{document}
\maketitle
\thispagestyle{myheadings}
\markright{{\em LA-UR-98-4640} \hspace{31mm} 
{\small accepted for publication in {\em Physical Review Letters} } \hfill \hspace{1mm}}

A dominant feature of optimally and underdoped cuprates is the
appearance of a pseudo gap in the normal state excitation spectrum.
The microscopic mechanism which is responsible remains a mystery.
A number of scenarios for explaining the pseudo-gap
have been proposed (see Ref.\ \onlinecite{randeria} for a recent
review).
However no calculations of the consequences of a large applied
field for the pseudo-gap have been published.
The high magnetic field behavior of the pseudo-gap provides
additional experimental characterization of the pseudo-gap
which is crucial for differentiating between various pictures.

We report very high accuracy measurements of the magnetic field
dependence of the $^{63}$Cu spin lattice relaxation rate in near
optimally doped \ybcod.
Our measurements demonstrate, in sharp contrast with previous
work,\cite{borsa,carrettaprb,carrettainc,mitrovicaps,mitrovicpreprint}
that there is {\it no} magnetic field dependence to \cuti\ in \ybcod.
This result has three important ramifications.
Although the magnetic fields we apply
shift \tc\ down by as much as 8 K,
the onset of pseudo gap effects does not shift down in temperature.
Hence the pseudo gap is unrelated to superconducting fluctuations,
even in near-optimally doped \ybcod\ where the gap
behavior appears just above \tc.
The onset of the pseudo-gap is very rapid, clearly demonstrating that
its magnitude is temperature dependent, opening very rapidly near 110
K.
Finally, the absence of any field effect indicates a
relatively large energy scale for the gap mechanism.
If dynamical pairing correlations or pre-formed pairs are
involved, the length scales must be very short.

The $^{63}$Cu NMR spin lattice relaxation rate reveals
the spin part of pseudo-gap behavior, the ``spin gap."
In underdoped YBa$_2$Cu$_3$O$_{6.6}$,
 $^{63}\left( {T_1T} \right)^{-1}$ famously exhibits a broad maximum
in the vicinity of room temperature and then
decreases as T approaches $T_c$ from
above.
In optimally doped \ybcod\ (data shown in
Figure 1) the maximum occurs at $\approx 110K$, and commences a
quite steep descent as T is lowered towards $T_c=93K$, though the
magnitude and onset temperature of the effect
seem to have a significant dependence on doping
level even for samples with the same $T_c$.\cite{martindalephil}  The
steepness of the downturn of $^{63}\left( {T_1T} \right)^{-1}$
in \ybcod\ enables a sensitive measurement.

The core experimental finding of this work is presented in Figure 1,
which shows, for magnetic fields of 0, 8.8, and 14.8 Tesla, the
$^{63}$Cu spin lattice relaxation rate $^{63}\left( {T_1T} \right)^{-1}$
vs. temperature.
All data shown in Figure 1 are normal state measurements, with
temperatures greater than $T_c\left( H \right)$.

The results of Figure 1 contrast sharply with previous measurements.
For example,
Carretta {\em et al.\/}\cite{carrettaprb,carrettainc} have found that
increasing magnetic field from 0 to 6 Tesla results in a decrease in
$^{63}\left( {T_1T} \right)^{-1}$ by some 20$\%$, for temperatures just
above $T_c$. They ascribe the decrease to the field suppression of phase
sensitive Maki-Thompson effects and conclude that their findings support
s-wave pairing. Borsa {\em et al.\/}\cite{borsa} observe similar
behavior but suspect a field effect upon antiferromagnetic fluctuations
as the mechanism.
In contrast, Mitrovic {\em et al.\/}\cite{mitrovicaps,mitrovicpreprint}
have probed $^{63}\left( {T_1T} \right)^{-1}$ indirectly through effects
upon the T$_2$ 
\begin{figure}
\leavevmode
\epsfxsize=3.25in
\epsffile{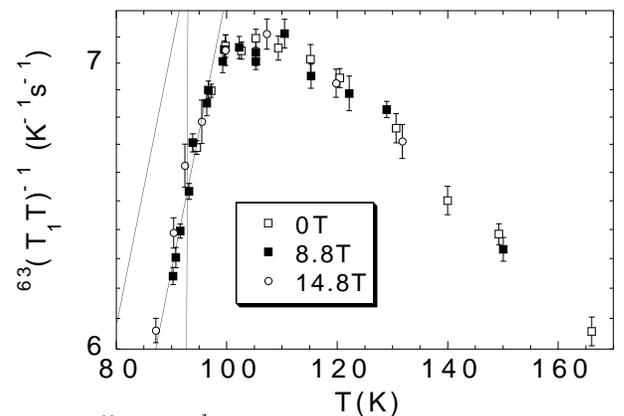}
\caption{$^{63}\left( {T_1T} \right)^{-1}$ vs. T for magnetic fields
(applied along the crystal c axis) of 0, 8.8, and 14.8 Tesla. Only the
data points such that T greater than $T_c$(H) are plotted. A straight
line is given which coincides approximately with the fall-off in
$^{63}\left( {T_1T} \right)^{-1}$, along with a parallel line shifted
to
lower temperature by an amount equal to 7.8K, the difference in $T_c$
for fields of 0 and 14.8 Tesla.
The vertical line indicates $T_c(H=0)$.} \label{fig:1overt1t}
\end{figure}
\noindent
of $^{17}$O\cite{recchiaprl}, and find that as the field
is increased from 2 to 24 Tesla the rate $^{63}\left( {T_1T}
\right)^{-1}$ {\em increases} by some 18$\%$.
Note that the field dependencies of 10-20$\%$ observed in all of these
previous measurements are approximately the same as the entire vertical
scale of Figure 1.

Now, we shall describe experimental procedures, focusing on possibly
significant differences with previous work.
Then we shall discuss inferences which can be drawn from the results.

Our measurements were carried out on powder samples aligned in epoxy
with the crystallite c axes parallel.
These samples were extensively characterized in earlier
studies\cite{mart:prb98}.
For all $T_1$ measurements the magnetic field is
applied along the sample c axis. The sample was prepared using the
procedure described in Ref. \onlinecite{Smithrpt}.
Figure 2 shows the high frequency $^{65}$Cu (3/2,1/2) satellite
transition lineshape, with a full width at half maximum of $\approx 400$
kHz. One important feature of Figure 2 is that the lineshape is not
symmetric-- there is essentially no intensity at frequencies much
greater than the line position of 136.7 MHz, but at lower frequencies
there is a significant ``background" intensity.
This kind of behavior for Cu NMR in aligned powder samples is well
understood: in a perfectly aligned powder sample there would be
background intensity neither above nor below the satellite transition.
The source of this background intensity, then, is the crystallites which
for whatever reason are not perfectly aligned.
Considering both the plane and chain $^{63,65}$Cu, and the full
interaction between the nucleus and the local electric field gradient
tensor, a misaligned crystallite in an 8.8T field can contribute NMR
intensity anywhere in the range from 69.1 to 136.7 MHz.
However, there can be $\em no$ intensity outside this range.
\begin{figure}
\leavevmode
\epsfxsize=3.25in
\epsffile{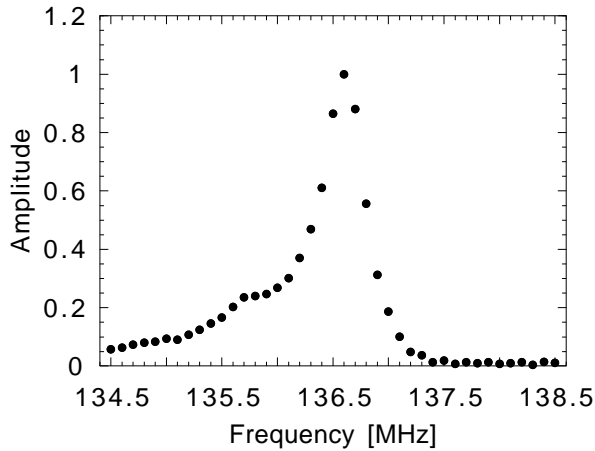}
\caption{
$^{65}$Cu NMR satellite (3/2$\leftrightarrow$1/2 transition)
lineshape with 8.8 Tesla magnetic field applied
along the c axis of the aligned powder sample.
No intensity is observed at frequencies significantly higher than the
peak frequency of 136.6 MHz, because that frequency is the highest
frequency which can occur for any Cu NMR transition and for any
orientation in YBa$_2$Cu$_3$O$_7$.
Thus, $T_1$ measurements performed on this transition are not corrupted
by background.}
\label{fig:Cu65sat}
\end{figure}
\noindent
Furthermore, intensity appearing at the upper (lower) frequency limit
derives exclusively from the $^{65}$Cu ($^{63}$Cu) upper (lower)
satellite transition with the field parallel to the c axis.

Conventionally Cu spin lattice relaxation has been measured on the
central (1/2,-1/2) transition, where one expects that background
intensity from misaligned crystallites will be present.
Martindale {\em et al.\/}\cite{Martindale94}, however,
have documented that in these
circumstances the background intensity can significantly contaminate
the signal  and reduce
the accuracy of the relaxation measurement.
For this reason, the $T_1$
measurements reported here have been performed on satellite
transitions
having the highest (or lowest) frequency which can be present. For the
14.8T measurements we used the low frequency $^{63}$Cu satellite. For
the 8.8T measurements we used the high frequency $^{65}$Cu satellite
of
Figure 2, and we plotted in Figure 1 the measured rate multiplied by
\( \left(^{63}\gamma / ^{65}\gamma \right )^2 = 0.8713 \).
Finally, for zero field we used the $^{63}$Cu nuclear
quadrupole resonance (NQR) transition at 31.5 MHz.

Figure 3 gives measured spin-lattice relaxation recovery curves at
T=100K,
demonstrating the importance, when making precise $T_1$ measurements,
of
probing at frequencies not subject to ``background" effects.
Experimental data are given for 0 T ($^{63}$Cu NQR),
and for both the high frequency $^{65}$Cu satellite
(3/2$\leftrightarrow$1/2) and the central (1/2$\leftrightarrow$-1/2)
$^{63}$Cu transition
at 8.8 Tesla.

Relaxation measurements were made by first inverting the
nuclear spin magnetization $M$
at time $t=0$. The recovery of $M(t)$ to its
thermal equilibrium value $M_{\infty}$ was then monitored,
using the CYCLOPS phase cycling sequence to remove
the effects of coil ringdown, gain 
\begin{figure}
\leavevmode
\epsfxsize=3.00in
\epsffile{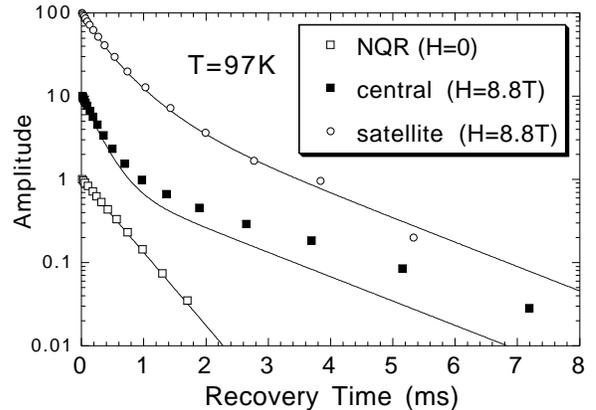}
\caption{$^{63}$Cu spin lattice relaxation experimental and
theoretical
recovery curve for zero field (NQR), and for H=8.8 Tesla, both central
transition and satellite transition. For the case of the satellite
transition data reported are for $^{65}$Cu, and the times are
multiplied
by $({}^{63}\gamma /{}^{65}\gamma )^2=0.8713$. Theoretical curves are
as
given in Table 1, using $T_1=1.484$ ms for all three sets of data.
The disagreement between the theoretical curve and experiment for
the case of the central transition demonstrates the error resulting
from background intensity, which we avoid in the data of Figure 1 by
observing the highest or lowest frequency satellite transitions.}
\label{fig:3recoverycurve}
\end{figure}
\noindent
imbalance,
and stimulated echoes\cite{Hoult}.
$^{63,65}$Cu is a four-level, spin 3/2 system,
and thus the relaxation curve  \( M(t)-M_{\infty}\)
is expected to be multi-exponential
with known coefficients\cite{narath}
but with only a single adjustable time constant
$T_1$ which is to be measured.
The expected functional forms for the relaxation of the
magnetization in the
three situations is given in Table 1.
In Figure 3 we use for all three relaxation curves a time constant $T_1$
equal to 1.484 ms, chosen to best fit the NQR data.
We see clearly from the figure that while the satellite and NQR
measurements follow the expected functional form beautifully, the
central transition has a significant deviation.
We interpret this effect as arising from ``background" intensity, having
a different $T_1$, at the frequency of the central transition. Such an
effect may be the source of the apparent field dependence observed in
previous measurements, with the exception of those of Refs.
\onlinecite{mitrovicaps,mitrovicpreprint}, which would not be
susceptible to this problem.
We note that the excellent, single exponential recoveries observed in
the NQR experiment rule out any effects due to spectral diffusion.

Now, what can be said about the field dependence of spin gap behavior
from Figure 1?
For illustration we have included in Figure 1 a straight line which
coincides approximately with the fall-off in $^{63}\left( {T_1T}
\right)^{-1}$ between $\approx 96K$ and $T_c$(H).
Then for comparison we include the same line, but shifted to lower
temperature by an amount equal to 7.8K, which is the expected difference
in $T_c$ for fields of 0 and 14.8 Tesla, assuming
dH$_{c2}$/dT=-1.9T/K.\cite{welp} One suspects that if the spin gap
phenomenon were a superconducting fluctuation effect, then the 14.8T
data would be shifted relative to the 0T data by a comparable amount,
but that plainly is not the case.
In fact, within experimental error the 14.8T data display the same onset
temperature as the 0T data, and from a close analysis of the data and
error bars we find that any decline in the temperature for onset of spin
gap effects over this range of fields must be less than 2K.
Thus, we find that application of a field (14.8T) representing a Zeeman
energy $g \mu_B H=20\,$K, which is $\approx 20\%$ of $T_c$, results in
no decrease in onset temperature within an uncertainty of 2\%.

Of course, a comparison of this result with quantitative predictions of
various theoretical models is necessary, but nevertheless we
suspect that this null result will pose a serious challenge to some
theories.
Possible exceptions include models\cite{leeprl96} based around the t-J
model, which call for local singlet pairing with an energy scale
governed 
\begin{table}
\caption{Functional forms of $T_1$ recovery curves expected for the zero
field NQR, and the high field central (1/2,-1/2) and satellite (3/2,1/2)
transitions, assuming a magnetic relaxation
mechanism\protect\cite{narath}.
\label{table1}}
\begin{tabular}{lc}

Transition & Recovery curve:
\( [M(t)-M_{\infty}]/[M_0-M_{\infty}] \) \\
\tableline \\  [-2ex]

NQR &
\(\exp \left( - \frac {3t} {T_1}    \right) \)  \\ [1ex]

central &
\(0.1 \exp \left( - \frac  {t} {T_1} \right) +
  0.9 \exp \left( - \frac {6t} {T_1} \right) \) \\[1ex]

satellite &
\(0.1 \exp \left( - \frac  {t}  {T_1} \right) +
  0.5 \exp \left( - \frac {3t}  {T_1} \right) +
  0.4 \exp \left( - \frac {6t}  {T_1} \right) \)

\end{tabular}
\end{table}
\noindent
by the exchange coupling J (of order 1000K).
We expect there would be a coupling of the applied magnetic
field to the orbital motion of pairs\cite{ek:sgprox} formed above \tc,
in this case the absence of a field effect will constrain the
length scale of the pair.
Gaps associated with the formation of ladder-like
structures\cite{sc:prl98} also involve large energy scales and so would
not be expected to be sensitive the fields applied here.  Finally the
antiferromagnetic Fermi liquid based approaches\cite{schmalprl} have
treated the pseudo-gap effect, but the extent of any magnetic
field dependence which would be predicted is not clear.

The fact that the onset temperature for spin gap effects
is not shifted down in temperature
along with the known suppression of $T_c$ demonstrates that the gap
even in optimally doped \ybcod\
is not closely tied to the onset of superconductivity,
and thus has nothing to do with superconducting fluctuations.
The abrupt decrease in \cuti\
then requires a strongly temperature dependent gap.
Clearly the gap is not present at high temperatures,
rather there must be an abrupt transition near 100K that
causes the gap to open.
\marginpar{*}
This abrupt opening could reflect the onset of charge ordering into
fluctuating structures which would enable the development of a
gap\cite{ek:sgprox,sc:prl98}.
Finally, the mechanism for the gap must have an energy scale
large compared to electron spin Zeeman energy scale $g\mu _BH$
($\approx 20K$ for a 14 Tesla field), even at optimal doping where the
gap only appears slightly above 100 K.

To conclude, we find that the spin gap effect in the cuprates is
insensitive to magnetic field, with the onset temperature
 remaining unchanged, within an uncertainty of 2K, by a 14.8T field
which suppresses $T_c$ by 8K. These results would appear
to call for relatively large energy and short length scales
in a scenario involving dynamical pairing correlations or pre-formed
pairs. However, in order to make a more definite statement,
quantitative
predictions from alternative theories will be necessary.

We gratefully acknowledge helpful discussions with P. A. Lee, J.
Schmalian, R. Klemm, A. Varlamov, A. Abrikosov, D. G. Stroud, J.
Wilkins, and T. R. Lemberger. 
Work at the Ohio State University was supported by the National Science
Foundation Division of Materials Research Contract No. NSF/DMR9357600,
by the U. S. Department of Energy, Midwest Superconductivity Consortium,
under Contract No. DE-FG02-90ER45427.
We gratefully acknowledge the support of the In-House Research Program
at the National High Magnetic Field Laboratory and the National High
Magnetic Field Laboratory which is supported by the National Science
Foundation through Cooperative Agreement No.\ DMR95-27035 and by the
State of Florida.


\end{document}